\documentclass{article}
\usepackage{cite}
\usepackage{amsmath,amssymb,amsfonts}
\usepackage{graphicx}
\usepackage{textcomp}
\usepackage{amsmath}
\usepackage{algorithm}
\usepackage{flushend}
\usepackage{authblk}
\usepackage{algpseudocode}
\usepackage[margin=1in]{geometry}
\usepackage{graphicx}
\usepackage{tikz}
\usepackage{multirow}
\usepackage{booktabs}  
\usepackage{siunitx} 
\usetikzlibrary{shapes.arrows}  
\usepackage{colortbl}
\definecolor{mygreen}{rgb}{1,1,1}
\definecolor{mygray}{rgb}{1,1,1}
\definecolor{myred}{rgb}{1,1,1}
\def\BibTeX{{\rm B\kern-.05em{\sc i\kern-.025em b}\kern-.08em
    T\kern-.1667em\lower.7ex\hbox{E}\kern-.125emX}}
\begin{document}
\title{Reducing Redundancy \\in Whole-Slide Image Patching \\for Scalable Indexing and Retrieval}
\author{Jialiang Geng, Ghazal Alabtah, Saghir Alfasly, Wataru Uegami, and H.R.Tizhoosh\\
KIMIA Lab, Dept. of Artificial Intelligence \& Informatics, \\Mayo Clinic, Rochester, MN, USA}

\maketitle

\begin{abstract}
The rapid growth of digital pathology has created an urgent need for efficient indexing and retrieval of whole slide images (WSIs). This need is intensified by emerging generative AI workflows, particularly retrieval-augmented generation (RAG), which require dependable similarity search to support high-stakes clinical decision-making. Yet the substantial cost of high-performance storage limits the scalability and accessibility of WSI indexing for many healthcare institutions. Consequently, methods that can reduce storage demands while preserving retrieval accuracy have become a critical research priority. We propose ARReST (Antithetical Redundancy Reduction Strategy), a principled oppositional framework that leverages redundancy across dissimilar tissue classes to markedly decrease the number of patches that must be indexed from each WSI. Instead of eliminating only within-class duplicates, ARReST identifies antithetical patches—those whose representations contribute minimally to cross-class discrimination—and prunes them from the searchable archive. This targeted reduction substantially compresses the index without sacrificing morphological diversity or retrieval fidelity. By minimizing superfluous patch representations, ARReST reduces storage footprint, lowers computational overhead, and accelerates similarity search across large pathology repositories. Extensive experiments on TCGA repository (The Cancer Genome Atlas with 21 organs) demonstrate that ARReST achieves significant index compression while maintaining competitive retrieval performance. The observed storage savings of 3\% to 60\% (14\%$\pm$13\%) can be reliably achieved without compromising retrieval performance for many organs. The proposed strategy enables scalable, cost-efficient WSI indexing and is well-suited for next-generation retrieval-driven clinical AI systems.
\end{abstract}

Whole Slide Image, Pathology, Patching, Image Retrieval, Redundancy 

\section{Introduction}
With the increasing number of whole slide images (WSIs) in digital pathology and the growing need for efficient search and retrieval of large WSI archives, effective indexing has become a critical necessity \cite{suresh2021optimization,tizhoosh2024image}. After tissue segmentation, patch selection is the first major step in many computational pathology algorithms.

Self-supervision is less suitable for patch selection because it forces models to learn surrogate tasks—such as predicting rotations or reconstructing masked regions—that do not directly align with identifying diagnostically meaningful or representative patches. Consequently, the selected patches may optimize performance on the pretext task rather than reflect true tissue relevance, leading to suboptimal indexing and retrieval.

Unsupervised methods are generally preferred in pathology because they eliminate the need for labor-intensive expert annotations, which are costly and often inconsistent across institutions. They also better capture the inherent variability of tissue without imposing predefined labels, enabling more generalizable and scalable representation learning.

Currently, all unsupervised patch selection techniques, such as mosaic \cite{Yottixel}, SPLICE \cite{Splice}, and SDM \cite{SDM}, adopt a ``divide'' strategy to manage the gigapixel scale of each WSI. These methods break down WSIs into a compact yet representative subset of patches to facilitate indexing and retrieval. However, significant bottlenecks and challenges remain. A practical WSI search engine and indexing strategy—with acceptable diagnostic accuracy—must balance storage requirements, cost efficiency, and the ability to perform rapid comparisons during both indexing and retrieval.

\emph{Democratization} of AI has become a major concern in recent years \cite{flotte2025democratizing,castro2024artificial}. In this context, storage costs and data management have emerged as critical challenges in digital pathology \cite{ardon2023digital,hanna2022integrating}. Several studies have highlighted the substantial expenses associated with handling and storing WSIs \cite{cost3,cost5}. The size of each WSI varies depending on magnification and resolution \cite{cost2,cost4}, typically ranging from 0.5 to 2 gigabytes per image. With the ongoing digitization of slides, it is estimated that maintaining an indexed archive—often referred to as an \emph{atlas}—may require 50–150 petabytes of high-performance storage annually \cite{Cost1}. Such requirements are highly inefficient and ultimately impede clinical deployment of large-scale digital pathology systems.

Current unsupervised methods extract a representative subset of patches from each WSI, reducing indexing storage to approximately 5–10\% of the original size. Although this represents a substantial improvement, the total storage demand—estimated at 5–15 petabytes—remains prohibitively large. Given that the cost per petabyte varies across institutions, the resulting financial burden on hospitals can be significant. Any further reduction in indexing storage would therefore be highly valuable, enabling broader adoption of digital pathology and more scalable retrieval solutions.

Beyond storage, the computational burden of comparing embeddings is another limitation of automated search systems. Most approaches compute distances between embeddings from the query WSI and those stored in the archive, typically requiring exhaustive pairwise comparisons. If an incoming WSI produces $m$ embeddings and another WSI contains $n$ embeddings, retrieval involves $m\times n$ pairwise distance calculations. Consequently, every reduction in the number of stored embeddings directly decreases the number of required comparisons by a factor of $n$, accelerating retrieval. Even when ``aggregation'' methods \cite{hemati2024short,chen2024benchmarking} are used to produce slide-level representations, redundancy reduction remains essential. Aggregation still depends on the underlying patch embeddings, and reducing their number lowers storage costs and accelerates computation without compromising downstream performance.

The primary contribution of this paper is the introduction of a similarity-matching procedure that removes redundancy by identifying highly similar patch-embedding pairs between WSIs of different classes. These pairs are stored in a redundancy repository and subsequently used to prune redundant embeddings from incoming WSIs before indexing or retrieval. This strategy not only reduces storage requirements but also significantly accelerates the comparison process during search.

The remainder of the paper is organized as follows. Section~\ref{sec:background} provides an overview of current search, retrieval, and indexing systems. Section~\ref{sec:method} describes the proposed similarity-matching approach in detail. Section~\ref{sec:results} presents experimental results on publicly available TCGA datasets.

\section{Background Knowledge}
\label{sec:background}

One of the major challenges in digital histopathology is the lack of efficient methods for extracting information from, and managing, the extremely large size of WSIs. Based on extensive benchmarking across public and private datasets using multiple performance metrics, the current baseline for WSI search and retrieval is the Yottixel engine \cite{Yottixel,lahr2024analysis}. The following subsections review the foundational components of the WSI processing pipeline—patch selection, retrieval, and indexing—used in many popular systems, including the Yottixel framework.

Given an atlas of WSIs with associated labels, slides are first divided into patches, after which a patch selection algorithm identifies a representative subset for each WSI. Embedding extraction and barcoding methods are then applied to enable fast retrieval. Incoming WSIs follow the same workflow, and their predicted labels are obtained by leveraging the stored embeddings and metadata of the atlas.

The method proposed in this paper, described in detail in Section~\ref{sec:method}, further reduces storage by shrinking the representation set of patches per WSI and decreases search time by removing redundant embeddings from incoming slides as well. Even when aggregation approaches are used to produce slide-level features, such redundancy reduction remains beneficial, since aggregation still depends on the underlying patch representations.

\subsection{Tissue Segmentation}
The first preprocessing step identifies tissue regions and removes non-informative areas such as pen marks or white background. An $n\times n$ grid is applied to each WSI, and tissue coverage is computed per grid cell. All cells containing less than 30\% tissue are discarded. Each remaining cell is treated as a patch of the WSI.

\subsection{Patch Selection Algorithms}
\begin{figure*}
    \centering
    \includegraphics[width=0.8\linewidth]{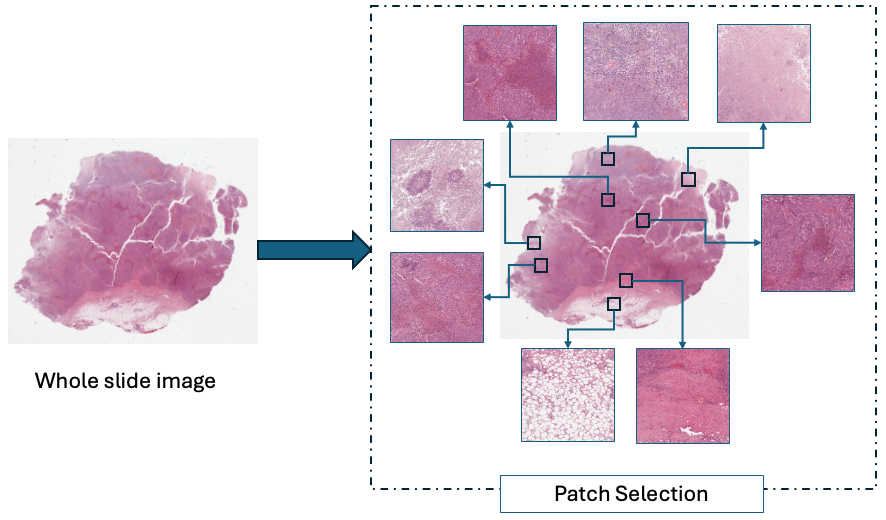}
    \caption{This schematic illustrations shows that patch selection methods select only extract a small number of sub-images from the gigapixel whole slide image of size 87,584 pixels by 72,452 pixels (Source: TCGA file TCGA-D8-A1XK-01Z-00-DX1.41EB1BBC-F230-4E3F-8C4E-CE331CAF1935.svs).}
    \label{fig:patchselect}
\end{figure*}

The most common strategy for extracting information from WSIs is to divide the slide into patches and select only a compact, representative subset. This step, known as \emph{patch selection}, is central to reducing computational and storage load. Numerous algorithms for this purpose have been introduced in recent years (Figure~\ref{fig:patchselect}).

\textbf{Yottixel} \cite{Yottixel} was previously considered the state-of-the-art patch selection algorithm. Its core idea is a two-step clustering strategy. The first step applies 
k-means clustering using color histograms, and the second step incorporates spatial information to refine sampling. With a fixed sampling rate within each cluster, the patch closest to the cluster centroid is selected as the representative.

\textbf{SPLICE} \cite{Splice} (Sequential Patching Lattice for Image Classification and Enquiry) also relies on color histograms but performs sequential sampling. After each selected patch, SPLICE removes similar patches from the sampling pool, thereby encouraging diversity and reducing redundancy in the selected subset.

\subsection{Foundation Model Embedding}
After selecting the representative patches from each WSI, we extract feature embeddings using a pre-trained or fine-tuned model—most commonly a foundation model. This step substantially reduces storage requirements and facilitates downstream processing such as BOB encoding, retrieval, and indexing. It also simplifies similarity computation, since comparisons are performed in a compact embedding space.

\subsection{Hashing and Barcoding}
Binary encoding is widely used to reduce storage demands and accelerate retrieval. The MinMax algorithm introduced in \cite{BOB} converts feature embeddings into binary codes by computing the sign of each one-dimensional deviation. This enables fast similarity analysis using Hamming distance instead of directly computing Euclidean distances between high-dimensional embeddings.

\subsection{Top-$k$ Majority Voting Prediction}
For label prediction of incoming WSIs, we retrieve information from the atlas embeddings using the same processing workflow. After computing BOB barcodes for the incoming slide, we calculate its distance to each atlas WSI, defined as the median of the minimum Hamming distances across all barcode pairs. We then identify the $k$ closest atlas WSIs and obtain their labels. The final predicted label for the incoming WSI is determined through majority voting among these 
$k$ nearest neighbors.

\section{Proposed Method}
\label{sec:method}
The main idea behind ARReST is based on the observation that when comparing representative patch embeddings from two WSIs belonging to different classes, patches originating from \emph{normal} tissue tend to exhibit higher pairwise similarity than patches representing abnormal or cancerous regions. This motivates a new redundancy-removal strategy: although labeled normal tissue is typically unavailable in WSI datasets, the availability of cancer-type labels enables the use of cross-class pairwise comparisons to identify and isolate highly similar (and therefore likely normal) patches. These patches can then be removed as redundant from each WSI.

The first step of ARReST is to perform similarity matching and construct a redundancy repository. Given labeled WSIs \((W_i, L_i)\), where \(W_i\) denotes the slide and \(L_i\) its class label, we apply a patch selection algorithm to extract a representative set of patches,
\[
\mathbf{P}_i = \{P_{ij}\}_{j = 1,2,\ldots,n_i},
\]
where \(P_{ij}\) is the \(j\)-th selected patch from slide \(W_i\), and \(n_i\) is the number of selected patches.

\begin{figure*}
    \centering
    \includegraphics[width=1\linewidth]{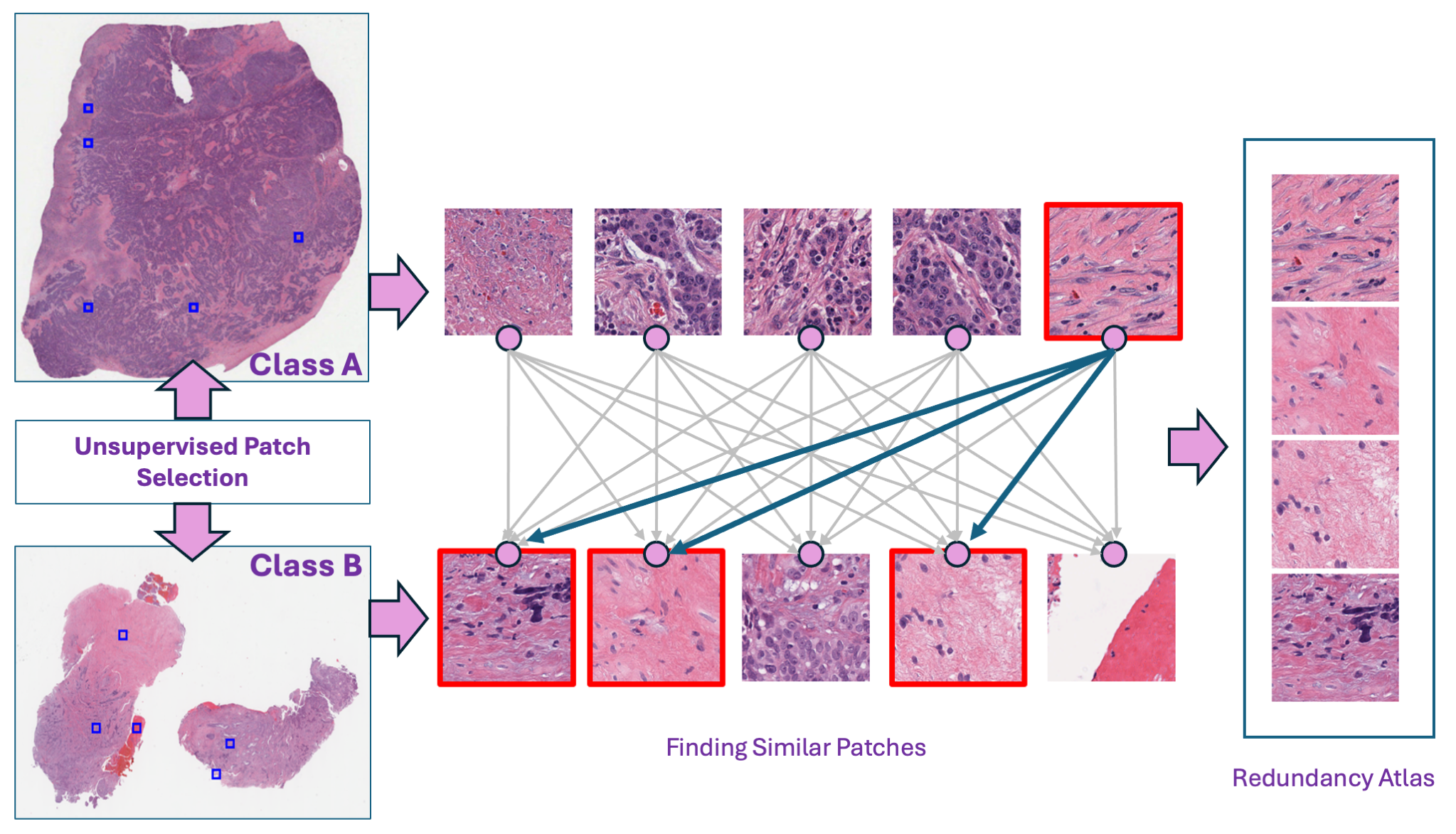}
    \caption{Workflow to establish an atlas of antithetically redundant patches. Two WSIs from different classes (e.g., two different cancer subtypes, hence the term \emph{antithetic}) are first selected. An unsupervised patch selection method is then applied to each slide to extract representative patches. These patches are compared pairwise to identify highly similar regions shared across the two classes. The resulting cross-class common patches—typically normal or non-discriminative tissue—are subsequently added to the Redundancy Atlas.
}
    \label{fig:red_repo}
\end{figure*}

Figure~\ref{fig:red_repo} illustrates the similarity-matching process (see Algorithm~\ref{alg:repo}). After patch selection (e.g., Yottixel, SPLICE, SDM), each patch is passed through a foundation model to obtain its embedding. We then repeatedly sample embedded patches from two WSIs of different classes and compute their pairwise similarity. Patch pairs with similarity scores below the \(q\)-quantile threshold of the global similarity score distribution are removed from both patch sets and stored separately in the redundancy repository.

For clarity, Figure~\ref{fig:red_repo} omits the embedding step and directly visualizes the removal of highly similar cross-class patches—typically corresponding to normal tissue—highlighting how redundancy is identified and eliminated within the ARReST framework.

\begin{algorithm}
\caption{Create redundancy repository}
\label{alg:repo}
\begin{algorithmic}[1]
\Require Patches embeddings \(\mathbf{P}_i=\{P_{ij}\}_{j=i,2,...n_i}\), labels \(L_i\),i=1,2,...,n. Similarity Score function f, prespecified quantile q.
\State \textbf{Initialize} Normal repository \(R\leftarrow\) [], Similar Score repository \(S\leftarrow\) []
\For{\(i\leftarrow\) 1 to n}
    \For{\(j \leftarrow \) 1 to n}
    \If{\(i\neq j\)}
    \For{\({P_{ik}} \in \mathbf{P}_i\)}
        \For{\(P_{jm} \in \mathbf{P}_j\)}
        \State Append \(f(P_{ik},P_{jm})\) to S 
        \EndFor
    \EndFor
    \EndIf
    \EndFor
\EndFor

\State get q quantile of S: \(S_q\)
\For{\(i\leftarrow\) 1 to n}
    \For{\(j \leftarrow \) 1 to n}
    \If{\(i\neq j\)}
    \For{\({P_{ik}} \in \mathbf{P}_i\)}
        \For{\(P_{jm} \in \mathbf{P}_j\)}
        \If{\(f(P_{ik},P_{jm})<S_q\)}
        \State Append \(P_{ik},P_{jm}\) to R
        \State Remove \(P_{ik},P_{jm}\) from \(\mathbf{P}_i,\mathbf{P_j}\)
        \EndIf
        \EndFor
    \EndFor
    \EndIf
    \EndFor
\EndFor

\State \textbf{return} Redundancy repository R, updated patch embeddings \(\hat{\mathbf{P}}_i\),i=1,2,...,n.
\end{algorithmic}
\end{algorithm}

Once the redundancy repository \(R\) is constructed and the atlas patch embeddings are updated to \((\hat{\mathbf{P}}_i, L_i)\), a common downstream step is to encode these embeddings as BOB barcodes \(\hat{\mathbf{B}}_i\). This conversion further reduces storage requirements and enables far more efficient distance computation, as Hamming distance between binary codes is significantly faster and simpler to compute than distances in continuous embedding space.

\begin{algorithm}
\caption{Remove redundancy and testing with top $k$ majority voting}
\label{alg:test}
\begin{algorithmic}[1]
\Require Incoming Patches embeddings \(\mathcal{P}=(\mathbf{P}_i)\) with \(\mathbf{P}_i=\{P_{ij}\}_{j=1,2,\dots,n_i}\). Similarity Score function f, q quantile of Similar Score repository S: \(S_q\), Redundancy repository \(R\), patches embeddings repository (as BOB barcodes) with labels and patients id \((\hat{\mathbf{B}}_i,L_i,G_i)\),i=1,2,\dots,n
\State \textbf{Initialize} 
\For{\(\mathbf{P}_i \in \mathcal{P}\)}
    \For{\({P_{ik}} \in \mathbf{P}_i\)}
    \For{Patch embedding \(R_j \in R\)}
        \If{\(f(P_{ik},R_j)<S_q\)}
        \State Remove \(P_{ik}\) from \(\mathbf{P}_i\)
        \State Break
        \EndIf
    \EndFor
    \EndFor
    \State Get the updated embeddings \(\tilde{\mathbf{P}}_i\)
    \State Get BOB barcode \(\tilde{\mathbf{B}}_i\) from updated embeddings \(\tilde{\mathbf{P}}_i\)
    \State Do majority vote of the labels \(L_{(1)},\dots,L_{(k)}\) from k closest BOB barcodes from repository \(\hat{\mathbf{B}}_{(1)},\dots,\hat{\mathbf{B}}_{(k)}\) which do not share the same patient id with \(\tilde{\mathbf{B}}_i\) and predict label \(\hat{L}_i\)
    
\EndFor

\State \textbf{return} Updated patch embeddings \(\tilde{\mathbf{P}}_i\), i=1,2,\dots,n and predicted labels \(\hat{L}_i\)
\end{algorithmic}
\end{algorithm}

Algorithm~\ref{alg:test} outlines the downstream processing and testing procedure. For each incoming WSI \(W_i\), we first apply a patch selection algorithm to obtain a representative set of patches and then use a foundation model to map these patches to embeddings \(\mathbf{P}_i \in \mathcal{P}\). For each patch embedding \(P_{ik} \in \mathbf{P}_i\), we compute its similarity score with every embedding \(R_j\) in the redundancy repository \(R\). If the similarity score falls below the \(q\)-quantile threshold of the similarity-score distribution \(S_q\), the corresponding patch embedding is removed from the representation set. This yields the updated embedding set \(\tilde{\mathbf{P}}_i\), which is subsequently converted into its BOB barcode representation \(\tilde{\mathbf{B}}_i\). We then use the atlas BOB barcodes, along with their associated labels and patient identifiers \((\hat{\mathbf{B}}_i, L_i, G_i)\), to predict the label of the incoming WSI via top-\(k\) majority voting.

\begin{figure}[H]
    \centering
    \label{fig:str}
    \includegraphics[width=1\linewidth]{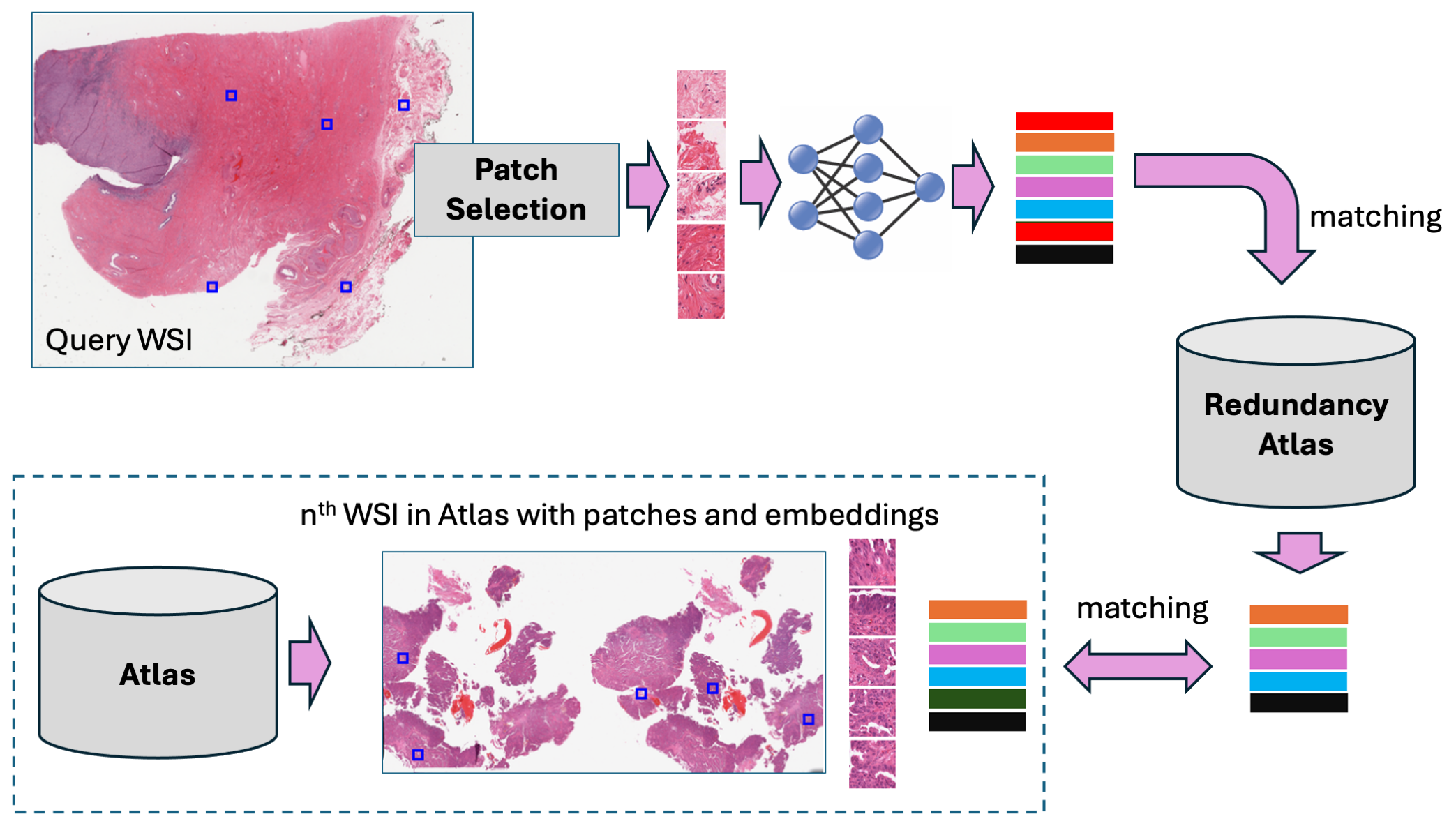}
    \caption{The structure of the search engine, when using an Atlas enhanced with a Redundancy Atlas, eliminates irrelevant patches after matching patch embeddings.}
    \label{fig:enter-label}
\end{figure}

Figure~\ref{fig:str} illustrates the complete pipeline of the proposed search engine and indexing framework.

\section{Results}
\label{sec:results}

In this section, we compare the search performance of the extracted embeddings before and after redundancy removal via similarity matching. We use Yottixel as the baseline patch selection algorithm, with WSIs processed at \(20\times\) magnification and a patch size of \(224 \times 224\). For deep feature extraction, we employ the UNI model, a publicly available pretrained vision encoder for histopathology (100M images, 100K WSIs) developed on diverse neoplastic, infectious, inflammatory, and normal tissues \cite{UNI}.

We evaluate performance using 5-fold cross-validation under both conditions. In each fold, evaluation is performed in a ``leave-patient-out'' manner to ensure that WSIs from the same patient never appear in both the test set and the atlas. This requires all slides from a given patient to be assigned to a single fold, while preserving the label distribution across folds. To satisfy these requirements, we implement Stratified Group \(K\)-Fold cross-validation. For each fold, the held-out set serves as the test dataset and the remaining folds form the atlas.

Each test WSI is patched, embedded, and compared against all embeddings in the redundancy repository to compute similarity scores. Embeddings with similarity scores below the quantile threshold are removed. The remaining embeddings are converted into BOB barcodes, and distances are computed between these barcodes and those stored in the atlas. We then retrieve the Top-1, Top-3, and Top-5 most similar atlas WSIs and determine the predicted label (i.e., cancer subtype for each organ) using majority voting.

Our experiments use The Cancer Genome Atlas (TCGA) dataset \cite{TCGA}, consisting of 11,679 WSIs. We evaluate performance separately for each organ. Table~\ref{tab:results} reports the accuracy and macro-averaged F1 score before and after similarity matching for the Top-1, MV@3, and MV@5 retrieval scenarios.

\begin{table*}[ht]
\centering
\caption{Comparison of performance before and after redundancy removal. Higher accuracy values and F1 scores are highlighted in \textbf{bold}. The storage saving \(S\) indicates the percentage reduction in required storage after redundancy removal. MV@3: majority vote among the top-3 retrieved WSIs; MV@5: majority vote among the top-5 retrieved WSIs.); $S$: storage saving.}
\label{tab:results}
\tiny
\resizebox{\textwidth}{!}{
\begin{tabular}{l|cccccc||cccccc|c}
\toprule
& \multicolumn{6}{c|}{\textbf{Accuracy}} 
& \multicolumn{6}{c|}{\textbf{Macro Average F1 Score}} \\
& \multicolumn{2}{c}{\textbf{Top-1}}
& \multicolumn{2}{c}{\textbf{MV@3}}
& \multicolumn{2}{c|}{\textbf{MV@5}}
& \multicolumn{2}{c}{\textbf{Top-1}}
& \multicolumn{2}{c}{\textbf{MV@3}}
& \multicolumn{2}{c|}{\textbf{MV@5}} \\
\textbf{Datasets}
& Before & After
& Before & After 
& Before & After 
& Before & After 
& Before & After 
& Before & After 
& $S$\\
\midrule
\textbf{Cervix} 
 & \cellcolor{mygreen}52 & \cellcolor{mygreen}\textbf{53} & \cellcolor{mygray}60 & \cellcolor{mygray}60 & \cellcolor{mygreen}61 & \cellcolor{mygreen}\textbf{62}
 & \cellcolor{myred}\textbf{28} & \cellcolor{myred}27 & \cellcolor{myred}\textbf{30} & \cellcolor{myred}28 & \cellcolor{mygreen}25 & \cellcolor{mygreen}\textbf{28}
 & 24\%\\

\textbf{Lung}
 & \cellcolor{mygreen}63 & \cellcolor{mygreen}\textbf{64} & \cellcolor{mygray}66 & \cellcolor{mygray}66 & \cellcolor{mygreen}67 & \cellcolor{mygreen}\textbf{68}
 & \cellcolor{myred}\textbf{23} & \cellcolor{myred}22 & \cellcolor{myred}\textbf{24} & \cellcolor{myred}21 & \cellcolor{myred}\textbf{22} & \cellcolor{myred}20
 & 17\%\\

\textbf{Testicles}
 & \cellcolor{mygreen}51 & \cellcolor{mygreen}\textbf{56} & \cellcolor{mygreen}56 & \cellcolor{mygreen}\textbf{59} & \cellcolor{mygray}60 & \cellcolor{mygray}60
 & \cellcolor{mygreen}35 & \cellcolor{mygreen}\textbf{39} & \cellcolor{mygreen}39 & \cellcolor{mygreen}\textbf{41} & \cellcolor{myred}\textbf{44} & \cellcolor{myred}42
 & 33\%\\

\textbf{Colon}
 & \cellcolor{mygray}86 & \cellcolor{mygray}86 & \cellcolor{mygreen}86 & \cellcolor{mygreen}\textbf{87} & \cellcolor{mygray}86 & \cellcolor{mygray}86
 & \cellcolor{mygreen}54 & \cellcolor{mygreen}\textbf{55} & \cellcolor{mygreen}57 & \cellcolor{mygreen}\textbf{59} & \cellcolor{mygreen}54 & \cellcolor{mygreen}\textbf{55}
 &6\%\\

\textbf{Esophagus}
 & \cellcolor{mygreen}91 & \cellcolor{mygreen}\textbf{92} & \cellcolor{mygreen}90 & \cellcolor{mygreen}\textbf{92} & \cellcolor{myred}\textbf{91} & \cellcolor{myred}90
 & \cellcolor{mygray}62 & \cellcolor{mygray}62 & \cellcolor{mygray}62 & \cellcolor{mygray}62 & \cellcolor{mygray}62 & \cellcolor{mygray}62
  &8\%\\

\textbf{Eye}
 & \cellcolor{myred}\textbf{51} & \cellcolor{myred}49 & \cellcolor{myred}\textbf{51} & \cellcolor{myred}49 & \cellcolor{myred}\textbf{51} & \cellcolor{myred}48
 & \cellcolor{myred}\textbf{42} & \cellcolor{myred}41 & \cellcolor{myred}\textbf{45} & \cellcolor{myred}42 & \cellcolor{myred}\textbf{36} & \cellcolor{myred}32
  &5\% \\

\textbf{Pancreas}
 & \cellcolor{myred}\textbf{82} & \cellcolor{myred}81 & \cellcolor{mygreen}87 & \cellcolor{mygreen}\textbf{88} & \cellcolor{mygray}88 & \cellcolor{mygray}88
 & \cellcolor{mygreen}41 & \cellcolor{mygreen}\textbf{43} & \cellcolor{mygreen}47 & \cellcolor{mygreen}\textbf{49} & \cellcolor{mygreen}45 & \cellcolor{mygreen}\textbf{48}
  &12\% \\

 \textbf{Bladder}
 & \cellcolor{myred}\textbf{85} & \cellcolor{myred}83 & \cellcolor{mygray}87 & \cellcolor{mygray}87 & \cellcolor{mygreen}87 & \cellcolor{mygreen}\textbf{88}
 & \cellcolor{myred}\textbf{68} & \cellcolor{myred}65 & \cellcolor{mygray}69 & \cellcolor{mygray}69 & \cellcolor{mygray}70 & \cellcolor{mygray}70
  &3\% \\

 \textbf{Rectum}
 & \cellcolor{mygray}84 & \cellcolor{mygray}84 & \cellcolor{mygreen}85 & \cellcolor{mygreen}\textbf{86} &\cellcolor{mygreen} 87 & \cellcolor{mygreen}\textbf{88}
 & \cellcolor{mygreen}62 & \cellcolor{mygreen}\textbf{64} & \cellcolor{mygreen}57 & \cellcolor{mygreen}\textbf{59} & \cellcolor{mygreen}53 & \cellcolor{mygreen}\textbf{55}
  &15\% \\
 
 \textbf{Thymus}
 & \cellcolor{mygray}34 & \cellcolor{mygray}34 & \cellcolor{myred}\textbf{38} & \cellcolor{myred}36 & \cellcolor{myred}\textbf{40} & \cellcolor{myred}36
 & \cellcolor{mygreen}27 & \cellcolor{mygreen}\textbf{28} & \cellcolor{myred}\textbf{35} & \cellcolor{myred}33 & \cellcolor{myred}\textbf{35} & \cellcolor{myred}32
  &12\% \\

 \textbf{Stomach}
 & \cellcolor{mygray}44 & \cellcolor{mygray}44 & \cellcolor{mygray}49 & \cellcolor{mygray}49 & \cellcolor{myred}\textbf{50} & \cellcolor{myred}48
 & \cellcolor{mygreen}31 & \cellcolor{mygreen}\textbf{32} & \cellcolor{mygray}34 & \cellcolor{mygray}34 & \cellcolor{myred}\textbf{36} & \cellcolor{myred}34
  &7\% \\

 \textbf{Skin}
 & \cellcolor{mygreen}85 & \cellcolor{mygreen}\textbf{88} & \cellcolor{mygreen}90 & \cellcolor{mygreen}\textbf{91} & \cellcolor{mygray}91 & \cellcolor{mygray}91
 & \cellcolor{myred}\textbf{32} & \cellcolor{myred}20 & \cellcolor{myred}\textbf{31} & \cellcolor{myred}19 & \cellcolor{myred}\textbf{23} & \cellcolor{myred}19
  &12\% \\

 \textbf{Chest/abdom.}
 & \cellcolor{myred}\textbf{59} & \cellcolor{myred}57 & \cellcolor{mygray}67 & \cellcolor{mygray}67 & \cellcolor{myred}\textbf{70} & \cellcolor{myred}\textbf{}69
 & \cellcolor{myred}\textbf{34} & \cellcolor{myred}31 & \cellcolor{myred}\textbf{36} & \cellcolor{myred}34 & \cellcolor{myred}\textbf{36} & \cellcolor{myred}34
  &4\% \\
 
 \textbf{Thyroid gland}
 & \cellcolor{mygreen}71 & \cellcolor{mygreen}\textbf{72} & \cellcolor{myred}\textbf{78} & \cellcolor{myred}76 & \cellcolor{myred}\textbf{79} & \cellcolor{myred}77
 & \cellcolor{mygray}48 & \cellcolor{mygray}48 & \cellcolor{myred}\textbf{49} & \cellcolor{myred}43 & \cellcolor{myred}\textbf{48} & \cellcolor{myred}43
  &4\% \\

\textbf{Head neck}
 & \cellcolor{mygreen}82 & \cellcolor{mygreen}\textbf{84} & \cellcolor{mygray}85 & \cellcolor{mygray}85 & \cellcolor{mygreen}84 & \cellcolor{mygreen}\textbf{85}
 & \cellcolor{myred}\textbf{45} & \cellcolor{myred}37 & \cellcolor{myred}\textbf{39} & \cellcolor{myred}25 & \cellcolor{myred}\textbf{29} & \cellcolor{myred}25
  &16\% \\

\textbf{Soft Tissue}
 & \cellcolor{mygray}46 & \cellcolor{mygray}46 & \cellcolor{mygray}47 & \cellcolor{mygray}47 & \cellcolor{mygreen}46 & \cellcolor{mygreen}\textbf{47}
 & \cellcolor{myred}\textbf{44}
 & \cellcolor{myred}38 & \cellcolor{myred}\textbf{45} & \cellcolor{myred}39 & \cellcolor{myred}\textbf{42} & \cellcolor{myred}38
 &23\% \\
 
\textbf{Adrenal glands}
 & \cellcolor{mygreen}79 & \cellcolor{mygreen}\textbf{81} & \cellcolor{mygreen}79 & \cellcolor{mygreen}\textbf{82} & \cellcolor{mygreen}80 & \cellcolor{mygreen}\textbf{81}
 & \cellcolor{myred}\textbf{41} & \cellcolor{myred}34 & \cellcolor{myred}\textbf{39} & \cellcolor{myred}33 & \cellcolor{myred}\textbf{34} & \cellcolor{myred}29
  &60\% \\
 
 \textbf{Uterus}
 & \cellcolor{myred}\textbf{75} & \cellcolor{myred}74 & \cellcolor{myred}\textbf{76} & \cellcolor{myred}75 & \cellcolor{mygreen}74 & \cellcolor{mygreen}\textbf{75}
 & \cellcolor{myred}\textbf{53} & \cellcolor{myred}50 & \cellcolor{mygray}52 & \cellcolor{mygray}52 & \cellcolor{myred}\textbf{47} & \cellcolor{myred}46
  &9\% \\

 \textbf{Breast}
 & \cellcolor{mygreen}73 & \cellcolor{mygreen}\textbf{74} & \cellcolor{mygreen}79 & \cellcolor{mygreen}\textbf{80} & \cellcolor{myred}\textbf{81} &\cellcolor{myred}80
 & \cellcolor{myred}\textbf{27} & \cellcolor{myred}25 & \cellcolor{myred}\textbf{31} & \cellcolor{myred}29 & \cellcolor{myred}\textbf{31} & \cellcolor{myred}29
  &7\% \\

  \textbf{Kidney}
 & \cellcolor{myred}\textbf{92} & \cellcolor{myred}91 & \cellcolor{mygray}93 & \cellcolor{mygray}93 & \cellcolor{mygray}92 & \cellcolor{mygray}92
 &\cellcolor{myred} \textbf{79} & \cellcolor{myred}76 & \cellcolor{myred}\textbf{78} & \cellcolor{myred}76 & \cellcolor{myred}\textbf{72} &\cellcolor{myred} 69
  &10\% \\

  \textbf{Brain}
 & \cellcolor{mygray}63 & \cellcolor{mygray}63 & \cellcolor{mygray}65 & \cellcolor{mygray}65 & \cellcolor{mygreen}65 & \cellcolor{mygreen}\textbf{66}
 & \cellcolor{mygreen}41 & \cellcolor{mygreen}\textbf{42} & \cellcolor{mygray}44 & \cellcolor{mygray}44 & \cellcolor{mygreen}44 & \cellcolor{mygreen}\textbf{45}
  &13\% \\
\bottomrule
$\mu\pm\sigma$ & 69$\pm$17 & 69$\pm$17 & 72$\pm$17 & 72$\pm$17 & 73$\pm$16 & 73$\pm$17 & 44$\pm$15 & 42$\pm$15 & 45$\pm$14 & 42$\pm$16 & 42$\pm$14 & 41$\pm$15 & 
\\
\bottomrule
\end{tabular}
}
\end{table*}

\textbf{Analysis of Results --}
When measuring ``accuracy,'' results improved in 26 experiments after ARReST was applied, remained unchanged in 20 cases, and decreased in 17 experiments. However, the averages, standard deviations, and high \(p\)-values indicate that the before/after accuracy values are essentially equivalent. Because accuracy is typically reported for \emph{balanced} datasets, this suggests that the observed storage savings of \(14\% \pm 13\%\) can be reliably achieved without compromising retrieval performance in balanced settings.

In contrast, when measuring ``F1 Score,'' performance improved in only 16 experiments, remained unchanged in 9 cases, and decreased in 38 experiments. The averages, standard deviations, and \(p\)-values below 0.05 indicate that the before/after F1 values are statistically different. Since F1 score is generally used for \emph{imbalanced} datasets, this suggests that the same level of storage saving may come at the cost of reduced retrieval accuracy in imbalanced scenarios.

The \underline{organ-level analysis} shows that the effectiveness of ARReST depends strongly on the underlying anatomical and histopathological characteristics of each tissue type. ARReST performs most reliably in organs with \textbf{homogeneous morphology and clear inter-class separability}, such as the cervix, rectum, and brain. In these cases, both accuracy and F1 score remain high despite a \(14\% \pm 13\%\) reduction in index size, indicating that the pruned patches were largely redundant.

Several organs—including testis, colon, pancreas, skin, and kidney—exhibited \textbf{no loss in accuracy}, demonstrating that ARReST can safely reduce storage even when morphology is moderately variable but consistent within dominant subtypes. Likewise, esophagus and bladder maintained their F1 scores, showing resilience to pruning even under imbalanced subtype distributions.

In contrast, organs with \textbf{high intraclass heterogeneity}—such as lung, head and neck, soft tissue sarcomas, adrenal gland, and uterus—were more sensitive to pruning. These tissues contain diverse or rare subtypes that depend on a small number of discriminative patches; removing such patches can disproportionately impact retrieval in imbalanced datasets, leading to decreased F1 scores. Nevertheless, accuracy often remained stable in these organs, suggesting that pruning primarily affects minority classes.

Overall, ARReST is most effective in organs with \textbf{stable and distinctive histologic signatures} and remains viable in moderately heterogeneous tissues where performance is unaffected. For highly imbalanced or morphologically diverse organs, adaptive or subtype-aware pruning strategies may be needed to preserve minority-class retrieval performance while still achieving meaningful storage reduction (Table~\ref{tab:compare}).

\begin{table*}[t]
\centering
\caption{Organ Groups and Their Anatomical/Histopathological Characteristics Relevant to ARReST Performance when analyzing \textbf{MV@5}}
\begin{tabular}{p{3.5cm} p{0.1cm}  p{4.5cm} p{0.1cm} p{5.5cm}}
\textbf{Storage Saving}  & & \textbf{Organs} & & \textbf{Key Anatomical/Histopathological Features} \\ 
\hline
& & & & \\
Higher Accuracy \& F1 & 
& Cervix, Rectum, Brain &
& Homogeneous histology; strong inter-class separability; pruning removes mostly redundant patches. \\
& & & & \\
Accuracy stays the same &
& Testis, Colon, Pancreas, Skin, Kidney & 
& Moderate intraclass variability; dominant morphologies are consistent; clear subtype boundaries support pruning without accuracy loss. \\
& & & & \\
Higher Accuracy only  & 
& Lung, Bladder, Head \& Neck, Soft Tissue, Adrenal Gland, Uterus & 
& High intraclass heterogeneity; many rare subtypes; pruning can remove discriminative minority patches. \\
& & & & \\
F1 stays the same  & 
& Esophagus, Bladder &
& Subtype imbalance moderate rather than severe; preserved minority-class features allow stable F1 after pruning. \\
& & & &\\
Higher F1 only &
& Colon, Pancreas &
& Dominant subtype morphology stable; ARReST preserves key discriminative patterns despite imbalance. \\

\hline
\label{tab:compare}
\end{tabular}
\end{table*}

\section{Discussion and Conclusion}
In this study, we proposed a novel approach—Antithetical Redundancy Reduction (ARReST)—designed to improve the efficiency of indexing and retrieval in whole slide image (WSI) search engines for digital pathology. By exploiting redundancy among dissimilar tissue classes, ARReST substantially reduces storage requirements while maintaining, and in some cases improving, retrieval accuracy. Experiments on TCGA datasets demonstrate that ARReST effectively minimizes the number of stored patches without compromising overall search performance.

A key insight from this work is that a large portion of normal patches are inherently redundant and appear across WSIs regardless of cancer type. Eliminating these redundant patches early in the pipeline significantly reduces storage demands while preserving the discriminative content required for accurate retrieval.

Beyond storage reduction, ARReST also accelerates the retrieval process. With fewer patches to compare, search time decreases, making the system more practical for large-scale deployments. This improvement is particularly valuable in clinical contexts, where rapid and accurate WSI retrieval can support diagnostic workflows and decision-making.

Nonetheless, the method has limitations. The effectiveness of redundancy removal depends on the chosen similarity threshold, which may not be optimal for all organs or datasets. In tissues with high morphological variability, overly aggressive pruning could risk discarding informative patches. Future work could explore adaptive thresholding strategies or deep learning–based similarity estimation to refine redundancy detection.

In summary, ARReST offers a practical and effective strategy for improving WSI search engines by reducing redundant storage and accelerating retrieval. By addressing both storage constraints and computational efficiency, ARReST provides a scalable solution for managing increasingly large WSI archives. Future extensions may involve more adaptive similarity measures and applying the method to medical imaging domains beyond histopathology.

\section*{Ethics and COI}
This study exclusively utilized publicly available data from The Cancer Genome Atlas (TCGA), which is fully de-identified and compliant with all relevant ethical and regulatory standards. No human subjects were recruited, and no institutional review board (IRB) approval was required. The authors declare that they have no competing interests related to this work.

\end{document}